\documentclass[english,american]{llncs}
\usepackage[T1]{fontenc}
\usepackage[latin9]{inputenc}
\usepackage{textcomp}
\usepackage{amsmath}
\usepackage{graphicx}
\usepackage{babel}
\begin{document}

\title{DNA Self-Assembly and Computation Studied\\
 with a Coarse-grained Dynamic Bonded Model}

\titlerunning{DNA Self-Assembly and Computation}

\author{Carsten Svaneborg\inst{1}, Harold Fellermann\inst{1,}\inst{2}, and Steen Rasmussen\inst{1,}\inst{3}}

\authorrunning{C. Svaneborg, H. Fellermann, and S. Rasmussen}

\institute{Center for Fundamental Living Technology, Department of Physics, Chemistry
and Pharmacy, University of Southern Denmark, Campusvej 55, DK-5320 Odense,
Denmark, \email{science@zqex.dk}
\and
Complex Systems Lab, Barcelona Biomedical Research Park, Universitat Pompeu Fabra, Dr. Aiguad\'e 88, 08003 Barcelona, Spain, \email{harold@sdu.dk}
\and
Santa Fe Institute, 1399 Hyde Park Road, Santa Fe NM 87501, USA, \email{steen@sdu.dk}
}
\maketitle
\begin{abstract}
We study DNA self-assembly and DNA computation using a coarse-grained DNA model
within the directional dynamic bonding framework {[}C. Svaneborg, Comp. Phys. Comm. 183, 1793 (2012){]}.
In our model, a single nucleotide or domain is represented by a single interaction site.
Complementary sites can reversibly hybridize and dehybridize during a simulation. This
bond dynamics induces a dynamics of the angular and dihedral bonds, that model the
collective effects of chemical structure on the hybridization dynamics. We use the DNA
model to perform simulations of the self-assembly kinetics of DNA tetrahedra, an
icosahedron, as well as strand displacement operations used in DNA computation.
\end{abstract}

\section{Introduction}

Sequence specific hybridization of DNA single strands makes DNA molecules
a flexible programmable building block. By choosing the right sequences, DNA
self-assembly behavior can be programmed to produce well defined nano-structures.
In the pioneering work of Seeman et al., branched DNA constructs have been utilized 
to self-assemble into a variety of structures~\cite{Seeman1982,Seeman1991,Seeman1994,Winfree1998}.
With DNA origamis Rothemund invented a way to fold long DNA single strands into well defined planar structures by adding a large number of short stabilizing oligomer strands\cite{Rothemund2006}.
Later it was demonstrated how to let the planar origamis
self-assemble into 3D nano-structures such as a box~\cite{Andersen2008}.
Ever since the pioneering work of Adlemann in 1994~\cite{Adl:1994},
DNA has also been recognized as a massively parallel, versatile, and inexpensive
computing substrate. In order for such substrate to be of practical
interest, however, it is desirable that the computational framework
is scalable and that individual computational elements can be combined to
form circuits. Recently, a scalable approach to enzyme-free DNA computing
has been proposed where circuits consist of relatively short DNA strands
that communicate via strand displacement~\cite{See:06,Qia:2011}.

The Poland-Scheraga (PS) model has been
very successful in predicting thermal melting and renaturing of long DNA strands~\cite{poland1966phase,jost2009unified}. It describes a DNA double strand as a 1D lattice where
each base-pair is either hybridized or open. To each state is associated a free energy
that has a sequence specific contribution from nearest neighbor interactions~\cite{SantaLucia}
as well as a polymer contribution from the conformational entropy of internal bubbles and frayed
ends. Generalizations of the PS model exists, where the single strands are represented as semi-flexible polymers on a 3D lattice~\cite{EveraersKumarSimm2007,JostEveraers2010}. This provides a conceptual simplification since the polymer free energy contributions are given implicitly.
The Dauxois-Peyrard-Bishop\cite{peyrard2008modelling} (DPB) model represents a DNA
double strand as a 1D lattice, but each base-pair is described by a continuous base-pair
extension. The DPB model is defined by a Hamiltonian which includes a hybridization
potential and a harmonic term penalizing deviations between nearest neighbor extensions.

The chemical structure of short DNA oligomers can be studied with atomistic molecular dynamics
simulations such as Amber\cite{case2010amber,cheatham2000molecular} and Charmm\cite{CHARMM2009,mackerell2000development}. However, if we are interested in mesoscopic properties
of long DNA molecules, it is more effective to utilize coarse-grained simulation models.
Coarse-graining is the statistical mechanical process by which microscopic details are
systematically removed, producing an effective mesoscopic model~\cite {langowski2006polymer,de2011polymer}.
The major computational advantage of coarse-graining is that it allows us to focus our
computational resources on studying the structures and dynamics at the mesoscopic level. 

Coarse-grained models describe a nucleotide by a small number of effective interaction sites.
In the ``three site per nucleotide'' model of de Pablo and co-workers, three sites represent
the phosphate backbone site, the sugar group, and the base, respectively\cite{Sambriski2009,sambriski2009sequence}.
There is also a number of ``two site per nucleotide'' models, e.g. the model of Ouldridge
and co-workers~\cite{ouldridge2010dna,OulridgeLouisDoye2011}, where one site represents the
base and another site the backbone and the sugar ring. Savelyev and Papoian~\cite{SavelyevPapoian2011}
have formulated a ``one site per nucleotide'' model. As the number of interaction sites per
nucleotide is reduced, the chemical structure is progressively lost. In simulations of DNA
tagged nanoparticles, even more coarse-grained models are used. DNA molecules have been
modeled e.g. as semi-flexible polymers with attractive sites on each monomer~\cite{hsu2010theoretical},
or as a single sticky site that can be hybridized with free complementary free sticky sites~\cite{Martines-VeracoecheaPRL2011}.
While the chemical structure of DNA has been completely eliminated, these models still retain
the DNA sequence specific hybridization effects on nanoparticle self-assembly.

We are interested in studying the statistical mechanics of hybridizing DNA strands
and in particular the kinetics of DNA self-assembly and DNA computation using a DNA
model that is as coarse-grained as possible. We have implemented a general framework
allowing directional bonds to be reversibly formed and broken during molecular
dynamics simulations\cite{SvaneborgCPC2012}. Along with the bonds, the angular and
dihedral interactions required to model the residual effects of chemical structure
are also dynamically introduced and removed as dictated by the bond dynamics. This
framework allows us to simulate reversible hybridization of complementary beads and
chains built from such beads. In the present paper, we study a minimal dynamic bonding
DNA model. For simplicity, we assume that the binding energy, as well as the bond, angular, and dihedral
potentials are independent of sequence, and we have chosen a force field that produces 
a flat ladder-like structure in the double stranded state. Our motivation for these
choices are to minimize the number of parameters required to specify the DNA model.

Dynamic bonding DNA models combine ideas from most of the existing DNA models.
We regard them as dynamic generalizations of statistical mechanical theories
and simultaneously as simplifications of coarse-grained DNA models.
As in the PS model, a complementary base-pair can either be hybridized or
open. When a base-pair is hybridized, it is characterized by a continuous
hybridization potential as in the DPB model. Dynamic bonding DNA models can also
be regarded as off-lattice generalizations of the lattice PS model
\cite{EveraersKumarSimm2007}. Rather than trying to model chemical structure
with interaction sites as in the ``two and three site per nucleotide'' models~\cite{Sambriski2009,sambriski2009sequence,ouldridge2010dna,OulridgeLouisDoye2011} 
dynamic bonding DNA models use angular and dihedral interactions to model the residual
effects of local chemical structure. Dynamic bonded DNA double stands can reversibly
melt and reanneal, which is not possible with the ``one site per nucleotide model''
of Savelyev and Papoian~\cite{SavelyevPapoian2011}. Finally, as in the sticky
DNA models\cite{Martines-VeracoecheaPRL2011}, a single bead in a dynamic bonding
DNA model can equally well represent a domain. 

Sect.~\ref{sec:DNA-model} presents the dynamic bonding DNA model, which is used in Sect.~\ref{sec:Results} to study self-assembly of DNA constructs and DNA-computing constructs.
Sect.~\ref{sec:Conclusions} ends the article with a conclusions.

\section{Dynamic bonding DNA model\label{sec:DNA-model}}

In the present dynamic bonding DNA model, single stranded DNA (ssDNA) is represented
by a string of nucleotide beads connected by stiff springs representing directional
backbone bonds. Instead of using a four letter alphabet representing the ACGT nucleotides,
in the present paper we increase the alphabet maximally to avoid getting trapped in
transiently hybridized  states. Physically, this corresponds to assuming that each
bead represents a short sequence of nucleotides i.e. a domain, and that two
non-complementary beads or domains are unable to hybridize.
A novel feature of our DNA model is that it involves dynamic hybridization
bonds, which are introduced or removed between complementary interaction
sites or beads when they enter or exit the hybridization reaction radius.
Along with the bonds, we dynamically introduce or remove angular and dihedral
interactions in the chemical neighborhood of a hybridizing bead pair.
These interactions are introduced based on the local bond and bead type
pattern, and hence allows us to retain some effects of the local chemical
structure in coarse-grained models. We utilize bonds carrying directionality
to represent the 3'-5' backbone structure of DNA molecules. This allows us
to introduce dihedral interactions that can distinguish between parallel
and anti-parallel strand alignments. We have implemented this framework
in a modified version of the Large-scale Atomic/Molecular Massively Parallel
Simulator (LAMMPS)~\cite{Lammps,SvaneborgCPC2012}. 

The DNA model relies on two ingredients, a Langevin dynamic for propagating
a system in time and space, and a dynamic directional bonding scheme~\cite{SvaneborgCPC2012}
that propagates the chemical structure of the system. The force on bead $i$
is given by a Langevin equation

\[
{\bf F}_{i}=-\boldsymbol\nabla_{{\bf R}_{i}}U-\frac{m}{\Gamma}{\dot{\bf R}_{i}}+{\bf \xi}_{i}\quad\text{with}\quad U=U_\text{bond}+U_\text{angle}+U_\text{dihedral}+U_\text{pair}.
\]

Here, the first term denotes a conservative force derived from the
potential $U$. The second term is a velocity dependent friction,
and the third a stochastic driving force characterized by
$\langle{\bf \xi}_{i}(t){\bf \xi}_{j}(t')\rangle=k_{B}Tm/(\Gamma\Delta t)\delta_{ij}\delta(t-t')$.
The potential $U$ comprises four terms representing bond, angular, dihedral, and
non-bonded pair interactions, respectively. The friction and stochastic
driving force implicitly represents the effect of a solvent with a specified
friction and temperature. The Langevin dynamics is integrated using a Velocity
Verlet algorithm with a time step $\Delta t=0.001\tau_L$ and $\Gamma=2\tau_L$
using a customized version of LAMMPS~\cite{Lammps,SvaneborgCPC2012}.

Here and in the rest of the paper we use reduced units defined by
the Langevin dynamics and DNA model. The unit of energy is $\epsilon=k_{B}T$,
where we set Boltzmann's constant $k_{B}$ to unity. The bead-to-bead distance
along a single strand defines the unit of length $\sigma$ which correponds
to the rise distance of DNA. The mass is $m=1$ for all beads. A Langevin unit
of time is defined as $\tau_{L}=\sigma\sqrt{m/\epsilon}$. The diffusion
coefficient of DNA model strand is $D(n)=k_B T \Gamma/(m n)$ where $n$ is the
total number of beads. This an be can be equated with the DNA diffusion
coefficient of a particular experimental conditions to obtain a time mapping.
Extrapolating the data in Ref. \cite{TinlandMM1997} yields
$\tau_L \approx 1.6 \times 10^{-12}s$ for $n=20$.

\begin{figure}
\centering
\includegraphics[width=0.6\textwidth]{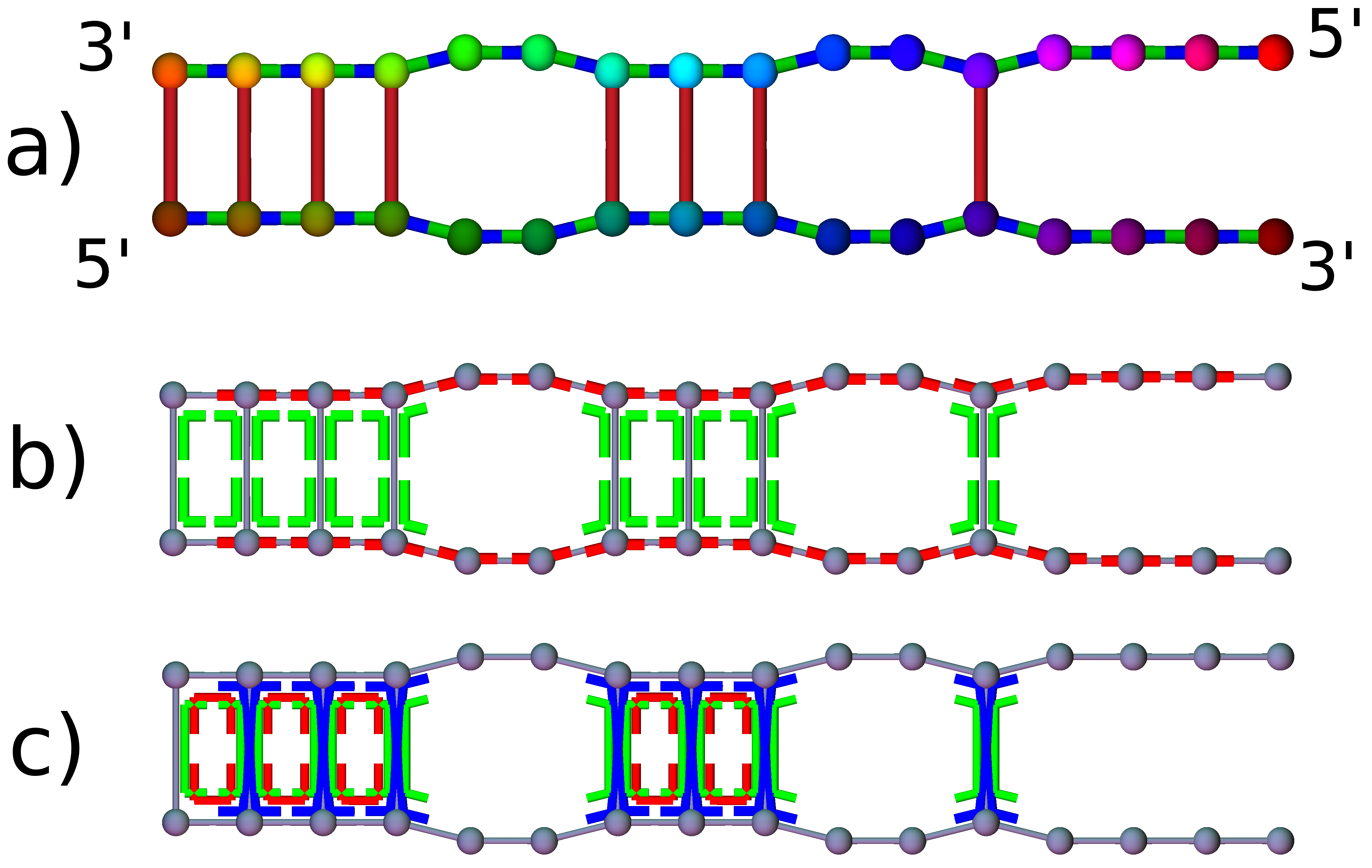}

\caption{Illustrative DNA conformation. a) complementary
beads, backbone and hybridization bonds, b) angular interactions indicated by two lines
parallel to the involved bonds, c) dihedral interactions indicated by
three lines parallel to the involved bonds. The figure is explained in
the text.}
\protect\label{fig:DNA-model}
\end{figure}

Fig.~\ref{fig:DNA-model}a shows complementary nucleotide beads with
the same hue but different levels of color saturation. As a simplification,
we allow each bead only to hybridize with a single complementary bead.
The DNA model has two types of bond interactions: permanent backbone
bonds (shown green/blue) and dynamic hybridization bonds (shown red).
Backbone bonds and hybridization bonds are characterized by the two potentials:

\[
U_\text{bond,bb}(r)=\frac{U_\text{min,bb}}{(r_\text{c}^\text{b}-r_{0}^\text{b})^{2}}\left((r-r_{0}^\text{b})^{2}-(r_\text{c}^\text{b}-r_{0}^\text{b})^{2}\right),
\]
and
\[
U_\text{bond,hyb}(r)=\begin{cases}
\frac{U_\text{min,hyb}}{(r_\text{c}^\text{h}-r_{0}^\text{h})^{2}}\left((r-r_{0}^\text{h})^{2}-(r_\text{c}^\text{h}-r_{0}^\text{h})^{-2}\right) & \mbox{for}\quad r<r_{c}^\text{h}\\
0 & \mbox{for}\quad r\geq r_\text{c}^\text{h}.
\end{cases}
\]

In the simulations, we use 
$U_\text{min,bb}=100\epsilon$, $r_{0}^\text{b}\equiv1\sigma$,
and $r_\text{c}^\text{b}=1.2\sigma$, $r_{0}^\text{h}=2\sigma$ and $r_\text{c}^\text{h}=2.2\sigma$.
Note that $U_\text{bond,hyb}(r)\leq0$ for all distances. When two non-hybridized beads of
complementary type are within a reaction distance $r_\text{c}^\text{h}$ a hybridization bond is
introduced between them. If they move further apart than $r_\text{c}^\text{h}$ again, the
hybridization bond is broken. The pair-interaction between beads is given by a soft
repulsive potential, while we use the same potential for angular and dihedral
interactions. They are given by
\[
U_\text{pair}(r)=A\left[1+\cos\left(\frac{\pi r}{r_\text{c}^\text{p}}\right)\right]\quad\mbox{for}\quad r<r_\text{c}^\text{p},
\]
where we use $A=1\epsilon$ and $r_\text{c}^\text{p}=1\sigma$ in the simulations, and
\[
U(\Theta;\Theta_{0},U_\text{min})=-\frac{U_\text{min}}{2}\left(\cos[\Theta-\Theta_{0}]+1\right),
\]

Along the backbone of single strands we use a permanent angular interaction defined by
$U(\Theta;\Theta_{0}=\pi,U_\text{min}=25\epsilon)$. This determines the persistence length
of single strands. In Fig.~\ref{fig:DNA-model}b backbone angular interactions are shown
as thick red lines around the central bead defining the angle.

In real DNA molecules, the hydrogen bonds between Watson-Crick complementary
nucleotides act together with stacking interactions and the phosphordiester
backbone bonds to give rise to a helical equilibrium structure of
the double strand. In our coarse-grained model, we utilize angular and dihedral
interactions to determine the ladder-like equilibrium structure of our DNA model.
To control the stiffness of the double strands  and to ensure anti-parallel
3'-5' alignment of the two single strands, we have assigned directionality
to the backbone bonds~\cite{SvaneborgCPC2012}. This is also necessitated
by the fact that the 3' and 5' carbons of the nucleotide sugar ring have been
merged into the single nucleotide bead. Fig.~\ref{fig:DNA-model}a
shows the backbone bonds colored green/blue to indicate the 3' and 5' ends,
respectively.

When a hybridization bond is introduced, we also dynamically add angular
interactions between the hybridization bond and the neighboring backbone
bonds. These angular interactions are characterized by the potential
$U(\Theta;\Theta_{0}=\pi/2,U_\text{min,a})$, which favors a right angle
conformation. When a hybridization bond is broken, concomitantly all the associated
angular interactions are removed. In Fig.~\ref{fig:DNA-model}b the angular
interactions are shown as green lines indicating the angle.

Besides introducing angular interactions, we also dynamically introduce
dihedral interactions. A dihedral interaction involves four beads
connected by three bonds, which defines a particular bond pattern,
where the bonds can either be a hybridization bond, a $3'-5'$ backbone
bond, or a $5'-3'$ backbone bond. Three bond patterns are possible.
The bond pattern corresponding to red dihedrals in Fig.~\ref{fig:DNA-model}c,
is characterized by $U(\Theta;\Theta_{0}=0,U_\text{min,d})$ which favors a
planar (cis) conformation. The bond pattern corresponding to blue dihedrals
is characterized by $U(\Theta;\Theta_{0}=\pi,U_\text{min,d},a=0)$ which favors parallel backbone
(trans) conformation. The last dihedral pattern corresponding to green
dihedrals is characterized by $U(\Theta;\Theta_{0}=0,U_\text{min,d})$ which favors
a parallel (cis) conformation. Note that without the directional backbone
bonds, we would not be able to distinguish between these two latter
dihedral patterns.

During a simulation, at each time we introduce a hybridization bond,
we also introduce up to four angular interactions and up to eight
dihedral interactions, less if the hybridization bond is at the end of
a strand. Let $\Delta$ be the total decrease in binding energy when 
two beads hybridize inside a chain, and we assign one third of this
energy to bond, angular, and dihedral interactions, respectively.
This choice does not affect the static properties of the model, which
are determined by the total energy associated with a conformation,
however it does influence the dynamic properties. Hence $U_\text{min,hyb}=\Delta/3$, $U_\text{min,a}=\Delta/12$, and
$U_\text{min,d}=\Delta/18$. We define
$\Delta=10\epsilon$ as a reference energy. Since only the ratio $\Delta/T$
enters the partition function of the model, this effectively fixes the
absolute melting temperature of the stands. With the present model, the
time spend per particle per step is approximately $1\times10^{-5}s$ on a standard PC.

\section{Results\label{sec:Results}}

Three dimensional DNA structures can be built by utilizing the self-assembly
properties of complementary strands and by linking several stands
into a e.g. end-linked constructs. In particular, we have designed
four constructs each comprising three end-linked 16 bead long strands.
By programming the complementarity of the strands, we have designed the
constructs to self-assemble into a tetrahedron\cite{Goodman09122005,Erben2007}.
We have also programmed the complementarity of 12 DNA constructs each
comprising 5 end-linked 8 bead long strands. These constructs have
been designed to self-assemble into an icosahedron\cite{icosahedra}.
We estimate that the melting temperatures are $T_{m}(8)\approx1.3\epsilon$,
and $T_{m}(16)\approx1.6\epsilon$ from a separate set of melting
simulations (not shown).

\begin{figure}
\centering
\includegraphics[width=0.2\columnwidth]{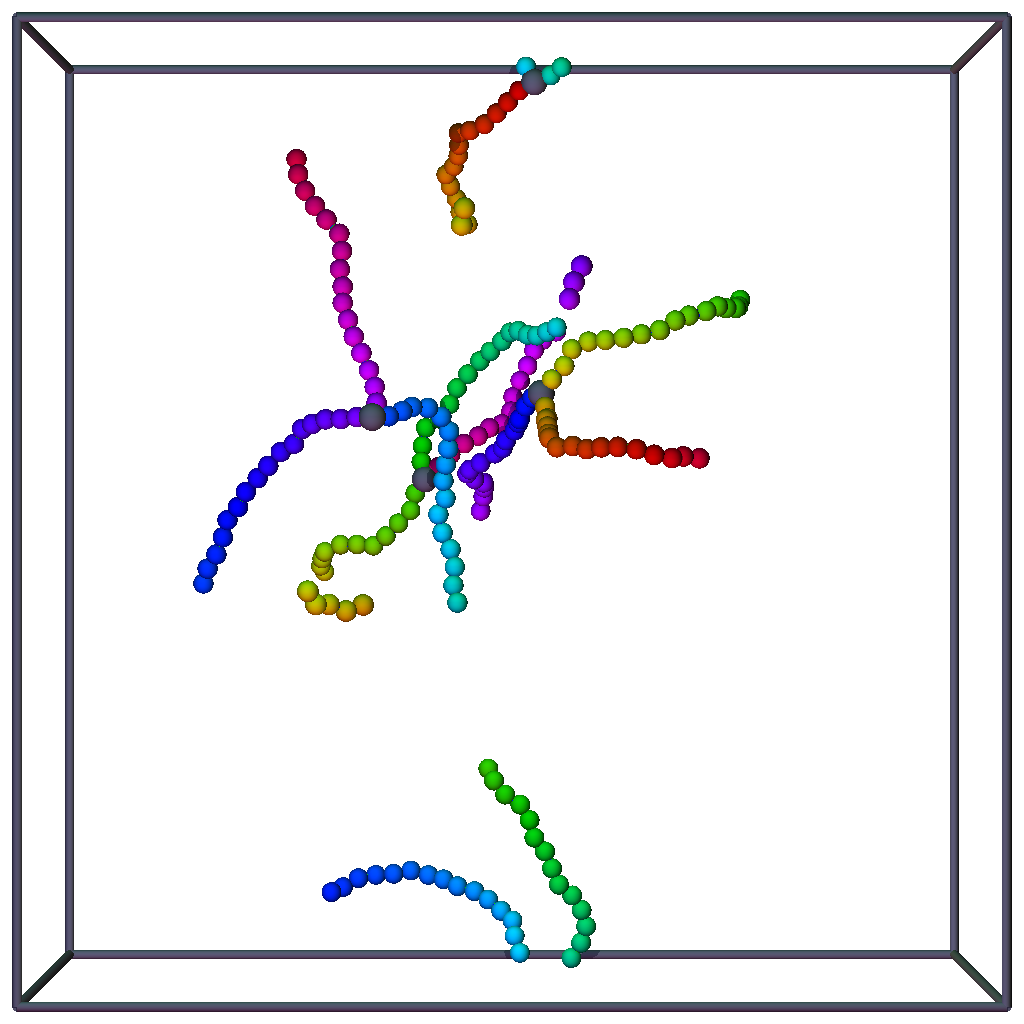}%
\includegraphics[width=0.2\columnwidth]{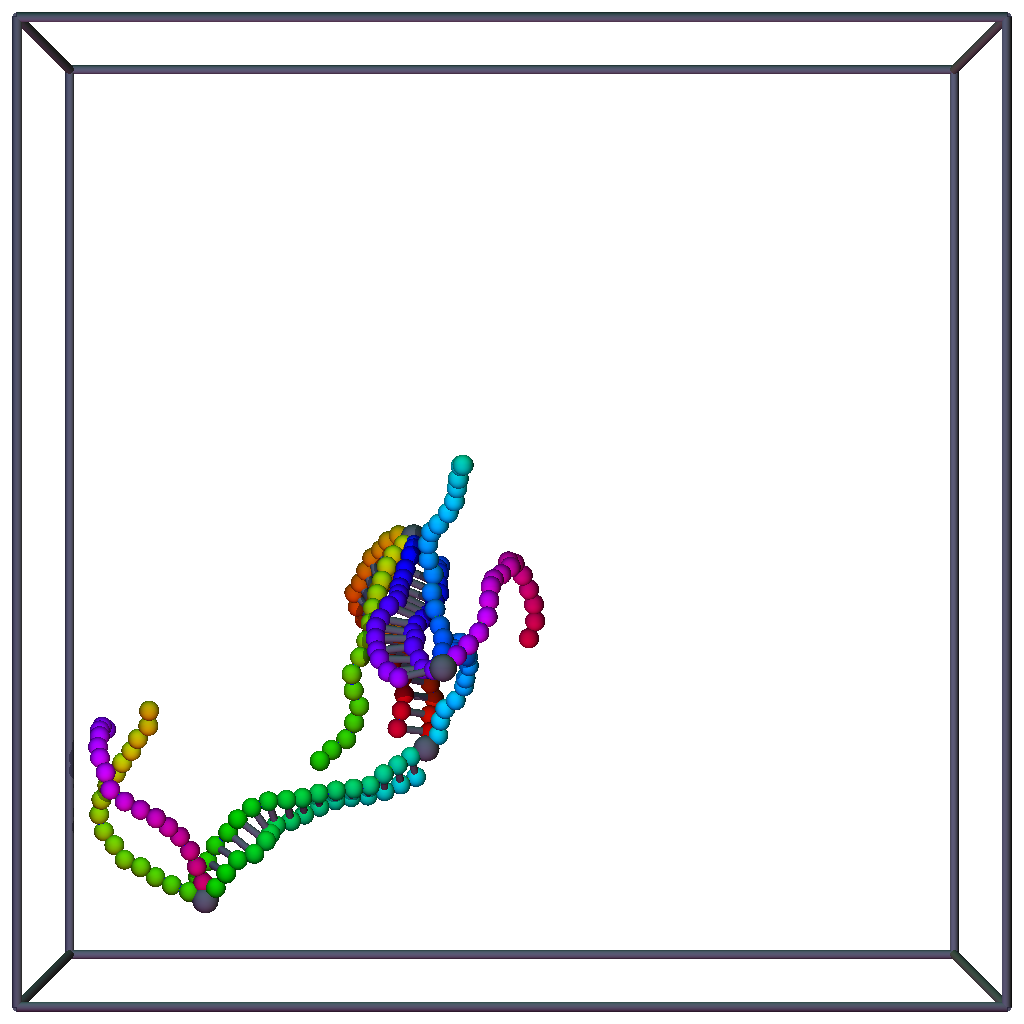}%
\includegraphics[width=0.2\columnwidth]{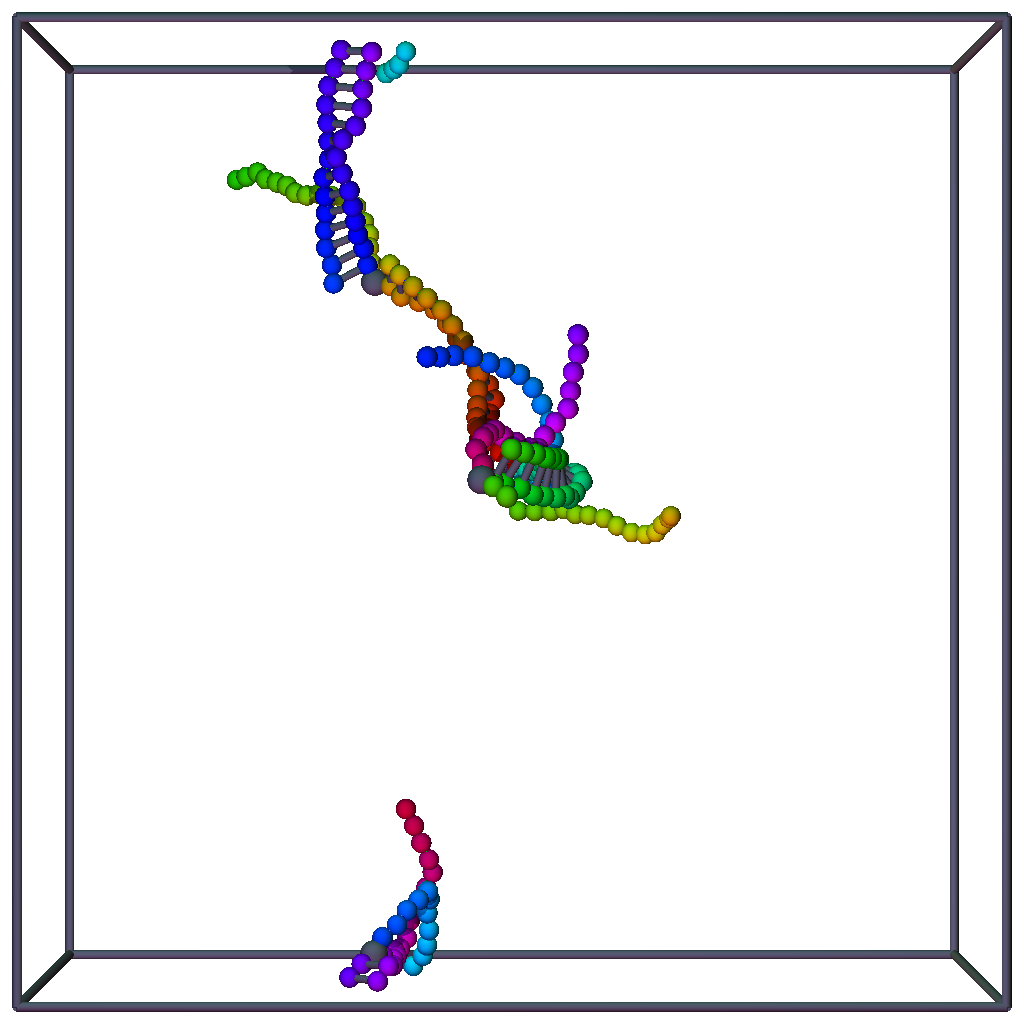}%
\includegraphics[width=0.2\columnwidth]{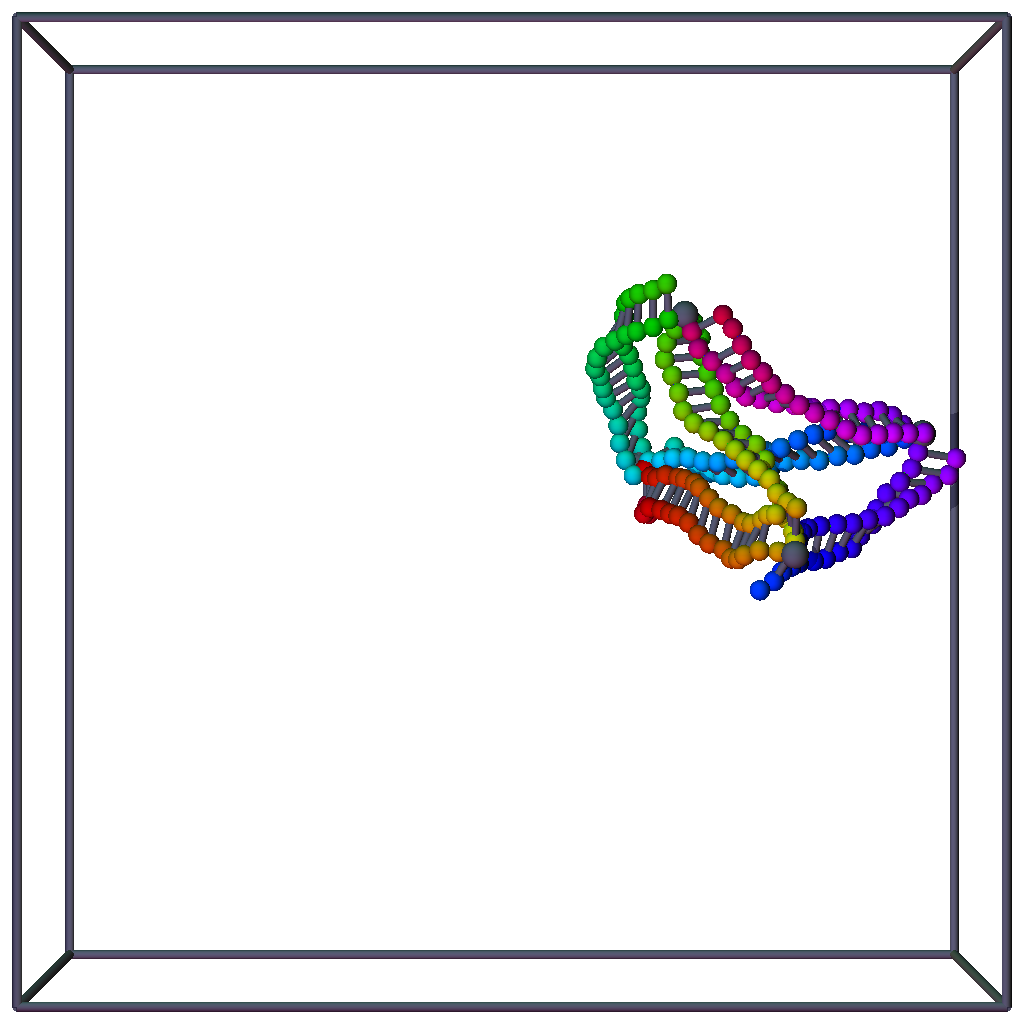}

\includegraphics[width=0.2\columnwidth]{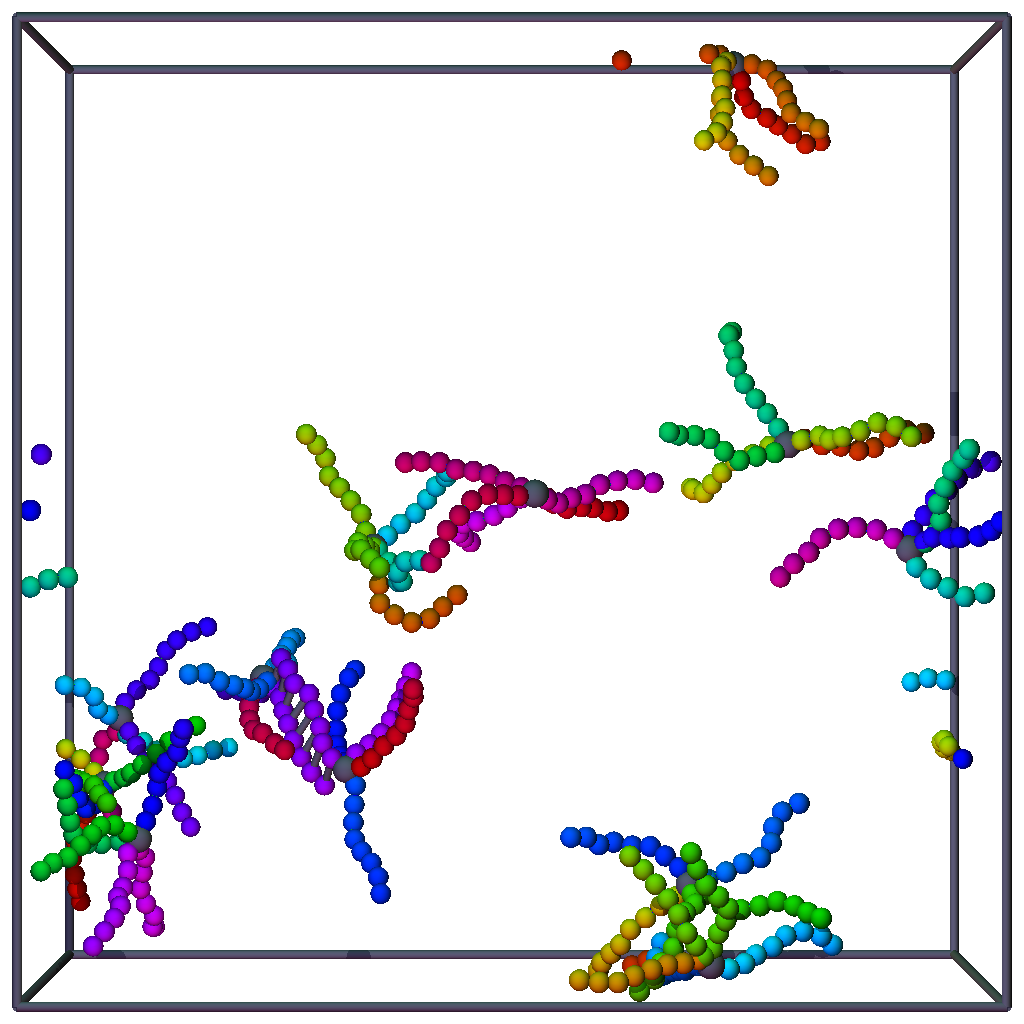}%
\includegraphics[width=0.2\columnwidth]{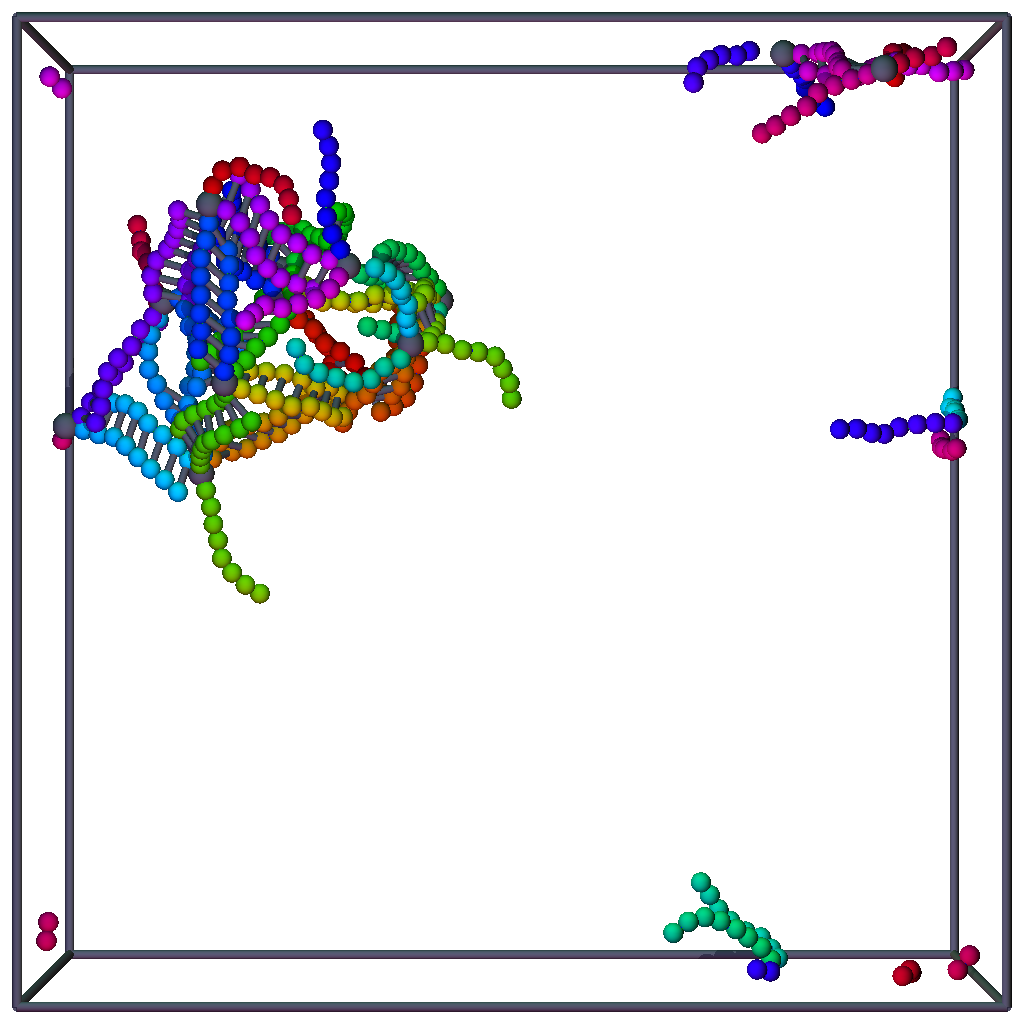}%
\includegraphics[width=0.2\columnwidth]{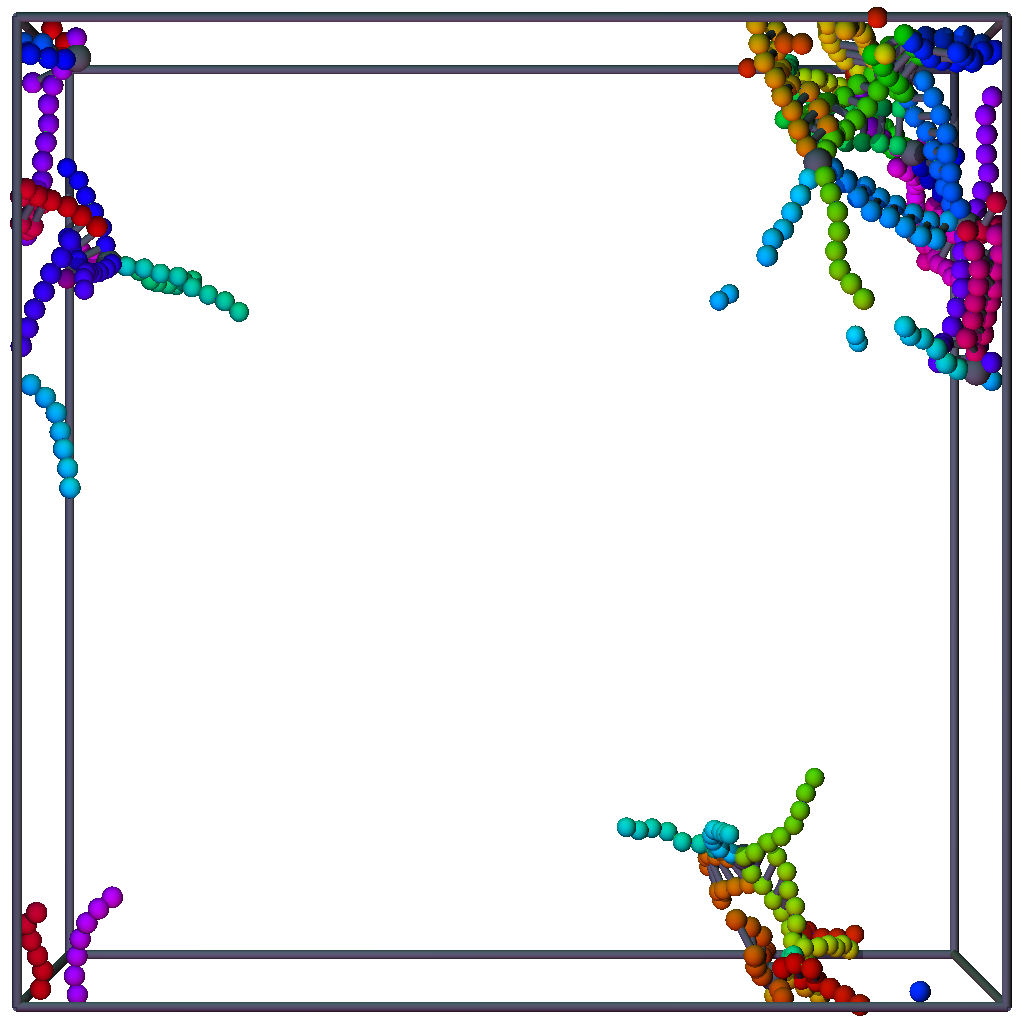}%
\includegraphics[width=0.2\columnwidth]{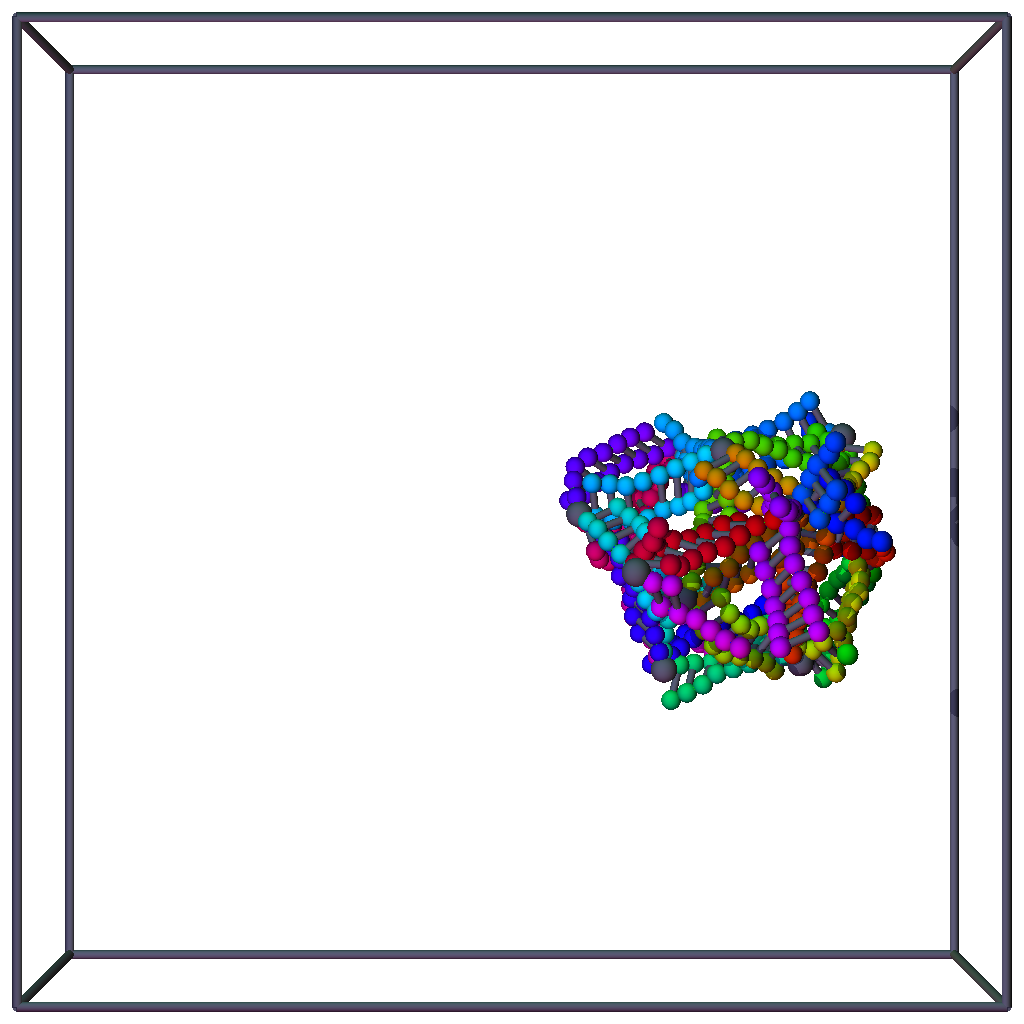}

\caption{\protect\label{fig:Self-assembly-of-tetrahedron}Self-assembly of a tetrahedron from four 3-functional DNA constructs
(top row) and an icosahedron from twelve 5-functional DNA constructs
(bottom row). Snapshots correspond to times $t=1000\tau_{L},$ $10.000\tau_{L}$,
$20.000\tau_{L}$, $50.000\tau_{L}$ steps (top row), and for $t=1000\tau_{L}$,
$20.000\tau_{L}$, $30.000\tau_{L}$, $60.000\tau_{L}$ steps (bottom
row). Simulations have been performed at $T=1.0\epsilon$.
Note that periodic boundary conditions apply to the simulation box.}
\end{figure}

\begin{figure}
\centering
\includegraphics[width=0.8\columnwidth]{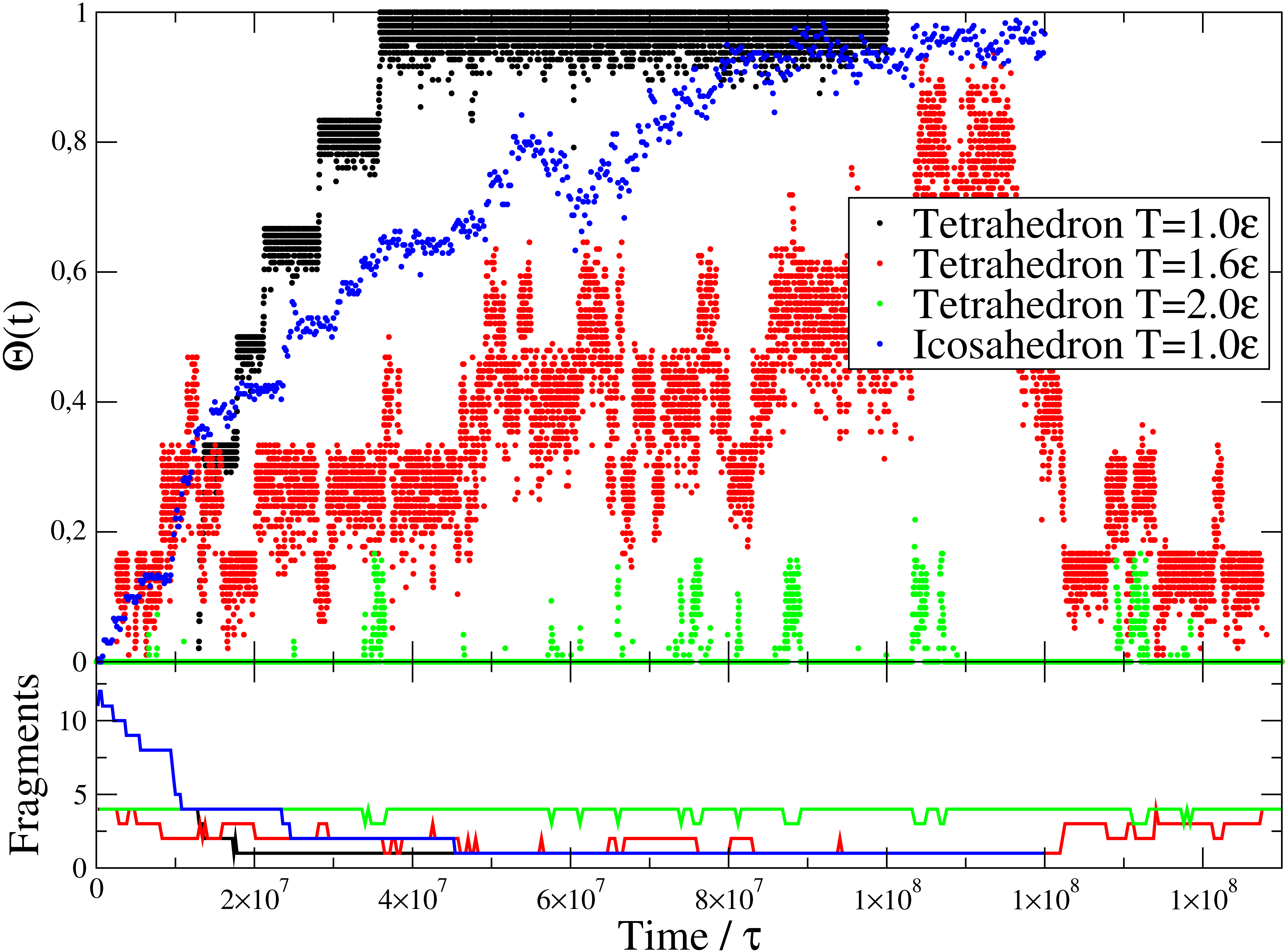}

\caption{\protect\label{fig:Hybridized-bond-fraction}Fraction of hybridized bonds (top) and number of fragments (bottom) vs. time and temperature during the self-assembly of a tetrahedron and an icosahedron.}
\end{figure}

Fig.~\ref{fig:Self-assembly-of-tetrahedron} shows visualizations
of the DNA constructs during the self-assembly process. Initially
the constructs are randomly placed into the simulation box. Progressively,
complementary strands hybridize with each other, and the constructs
form fragments that ultimately yield the designed target structures.
The time scale of the self-assembly dynamics is determined by the
time it takes the constructs to diffuse, collide, and hybridize completely.
Since we have the simulation trajectory, we can also characterize
the detailed time dependence of the self-assembly dynamics. Fig.~\ref{fig:Hybridized-bond-fraction}
shows the fraction of hybridized bonds as a function of time. By analyzing
the bond structure, we can furthermore study the evolution of the number of
structural fragments. For the icosahedron, we see a slow increase in
the hybridized bond fraction towards unity as the structure
is progressively assembled, while the number of fragments drops
simultaneously from initially twelve free constructs to one when all constructs
form a single icosahedron. However, even through we only have a single fragment,
it takes further time for the remaning hybridization bonds to be formed.
The equilibrium hybridization bond fraction
does not appear to be reached at the end of the simulation at $1\times10^{8}\tau$.
For the tetrahedron, we observe a similar increase in the fraction
of hybridized bonds, however with six distinct steps corresponding
to the hybridization of each edge of the tetrahedron.

The self-assembly dynamics is stochastic and depends on initial conditions
and random diffusive motion. We have run some of the simulations twice to
see how they approach equilibrium along different trajectories. The
equilibrium hybridization bond fraction appears to have been reached
by the tetrahedron self-assembly simulations. For tetrahedra, we observe
that self-assembly at higher temperatures leads to a marked decrease in
the average hybridization bond fraction similar to melting of DNA double
strands. At $T=1.8\epsilon$ the temperature is above the melting temperature
of the DNA constructs, and they only transiently hybridize. Since
we have a single fragment at equilibrium, the bond reduction is most
likely due to DNA bubbles. From the data sets we can estimate that
the melting temperature of the tetrahedron i.e. $\Theta(T_\text m)=0.5$ is
approximately $T_\text{m}(\mbox{Tetrahedron})\approx1.5\epsilon$.

In the strand displacement approach to DNA computation, individual gates consist of one DNA template that
is composed of several logical domains. In their initial state, all
domains but one are hybridized to one or more complementary strands
and are therefore inert. The only exposed single strand domain of
each gate is a short toehold region at one end of the template. This
toehold region can reversibly bind a complementary signal strand which
is designed to be longer than the toehold domain and complementary
to the next domain(s) of the template. The newly binding signal is
then able to hybridize to all matching domains of the template, thereby
displacing strands that where previously bound~\cite{Zha:2009}. The
displaced strands can be fluorescently marked output signals, or
internal signals that can bind to toehold regions of downstream gates.
By choosing domains of appropriate length, it can be guaranteed
that toehold binding is reversible, whereas the final strand displacement
is effectively irreversible, thus computation is energetically downhill
and kinetically irreversible, if and only if the correct input strands
are present and match the logical setup of the gates. It has been
shown that this approach leads to modular logic gates that enable
the design of large scale DNA circuits~\cite{Car:2011,Lak:2012}.

\begin{figure}
\centering
\includegraphics[width=0.2\columnwidth]{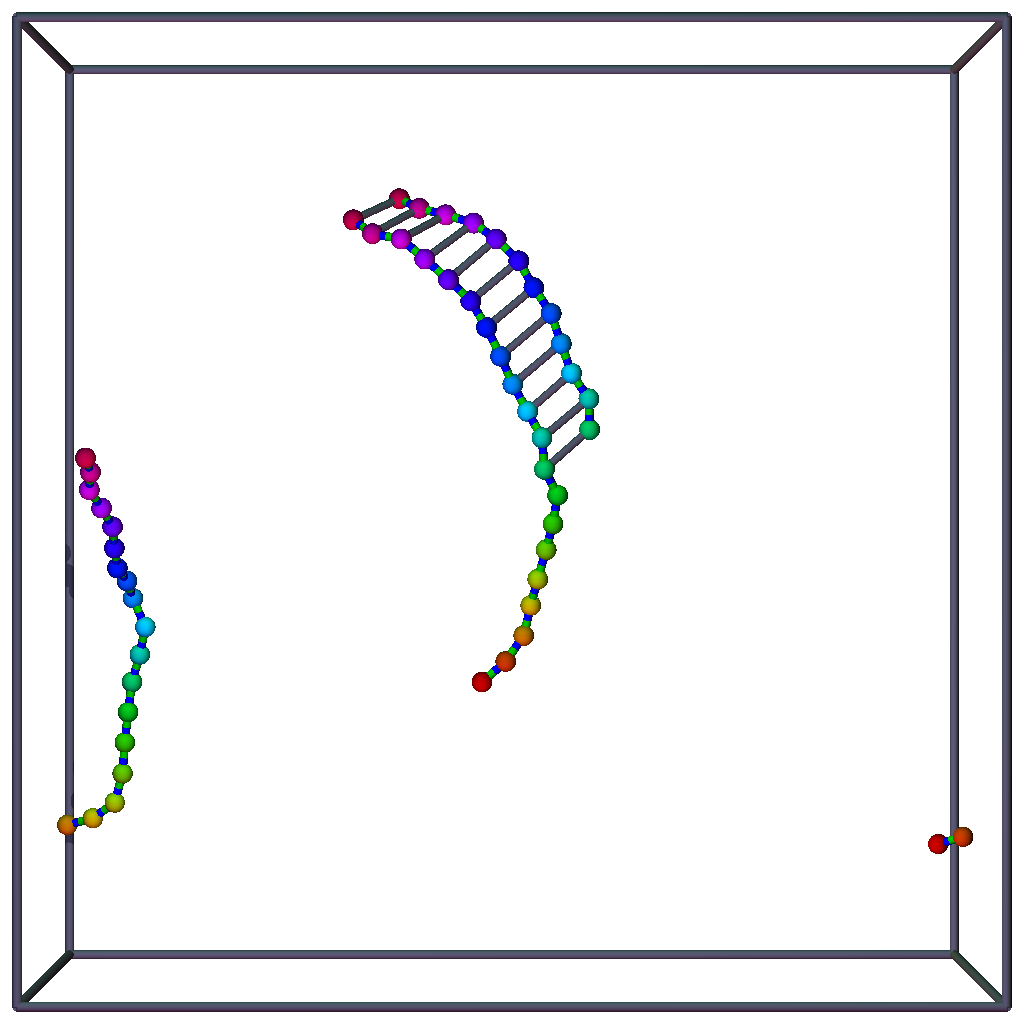}%
\includegraphics[width=0.2\columnwidth]{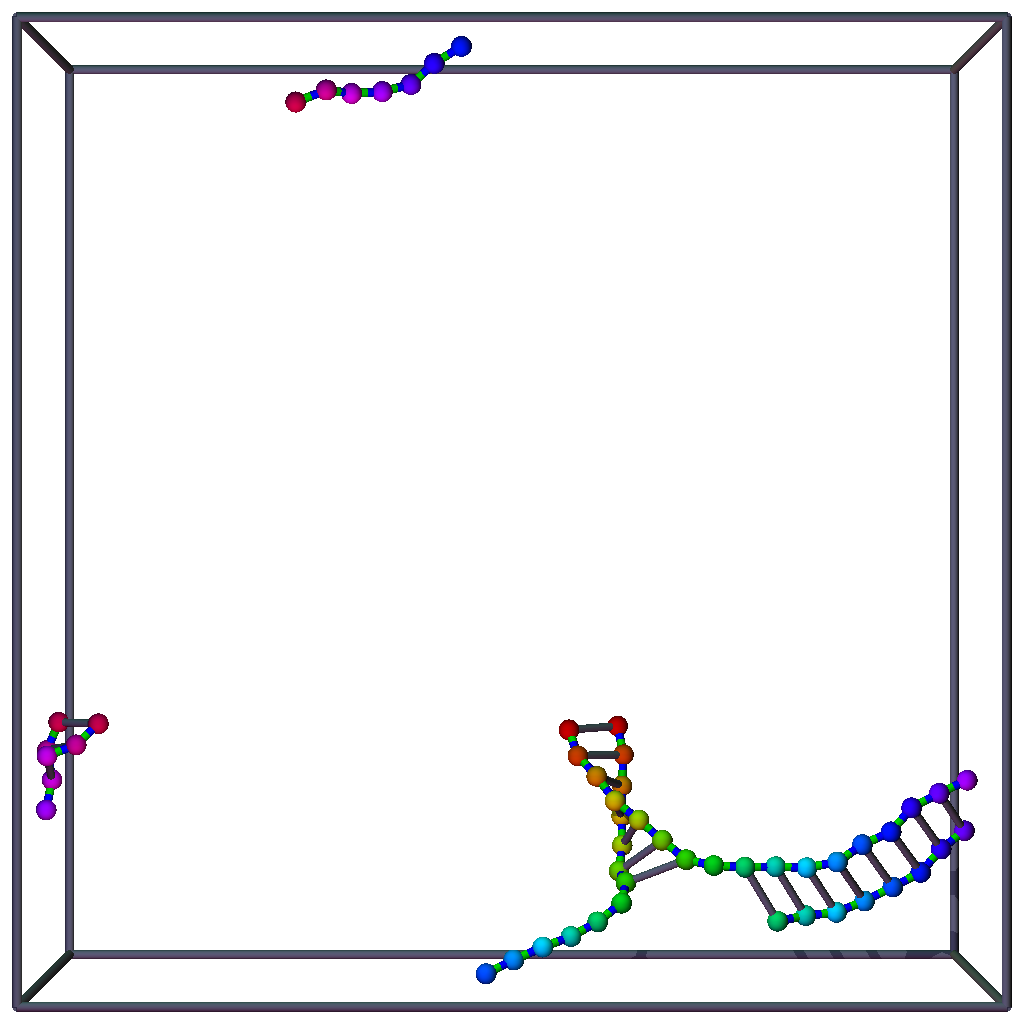}%
\includegraphics[width=0.2\columnwidth]{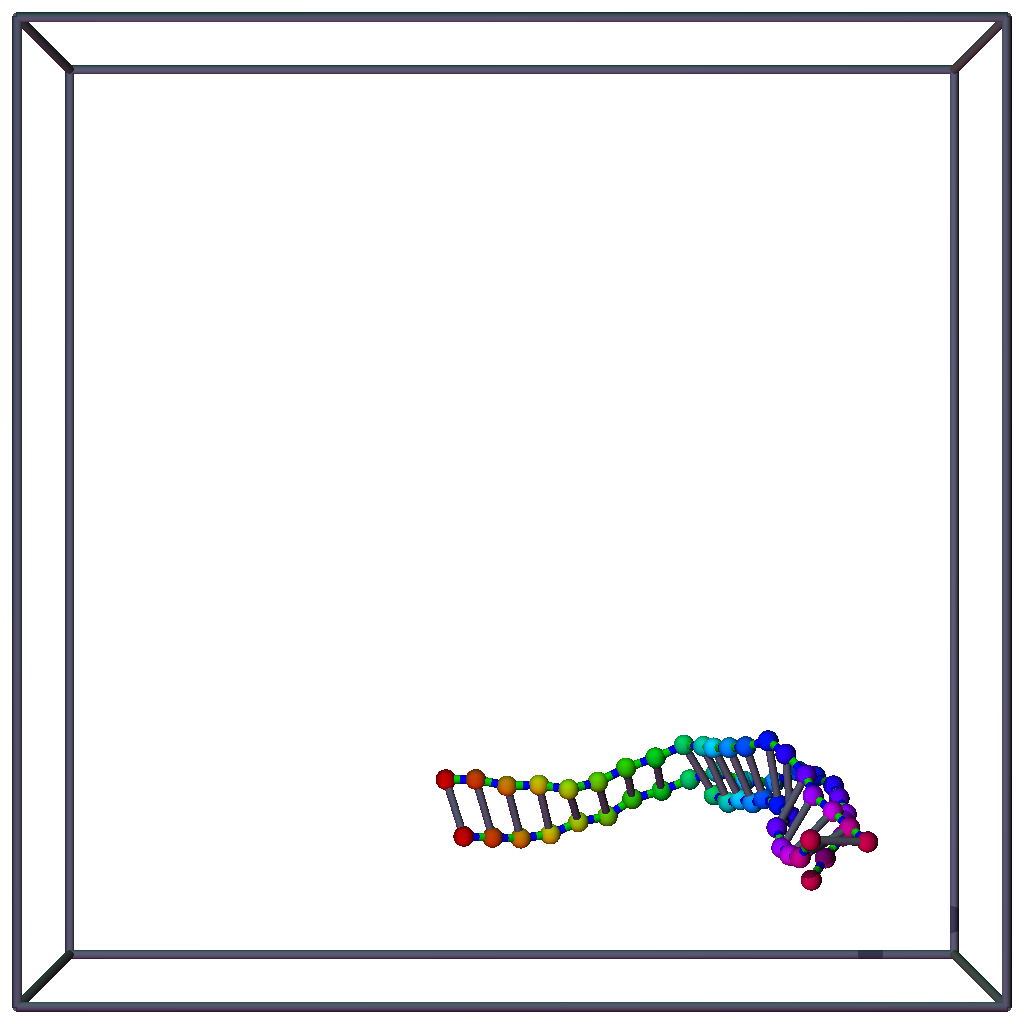}%
\includegraphics[width=0.2\columnwidth]{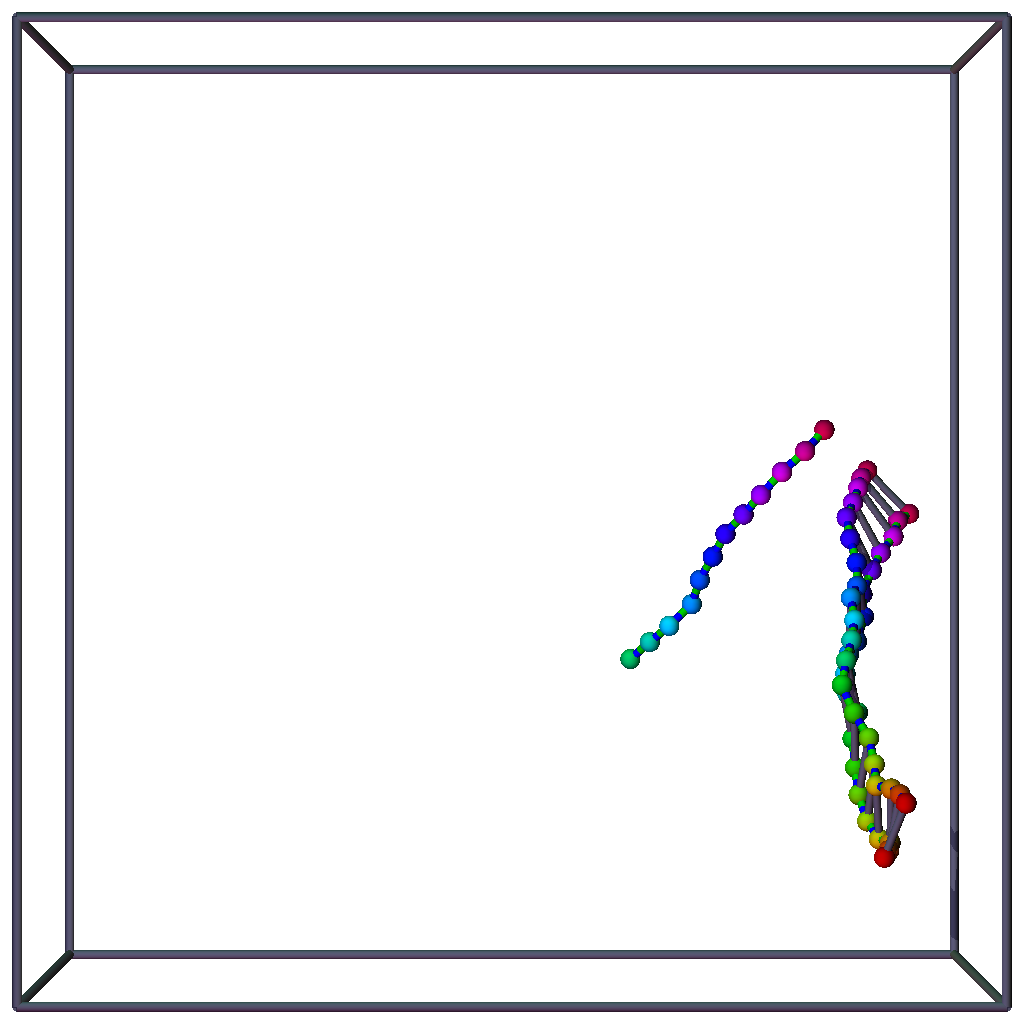}

\includegraphics[width=0.2\columnwidth]{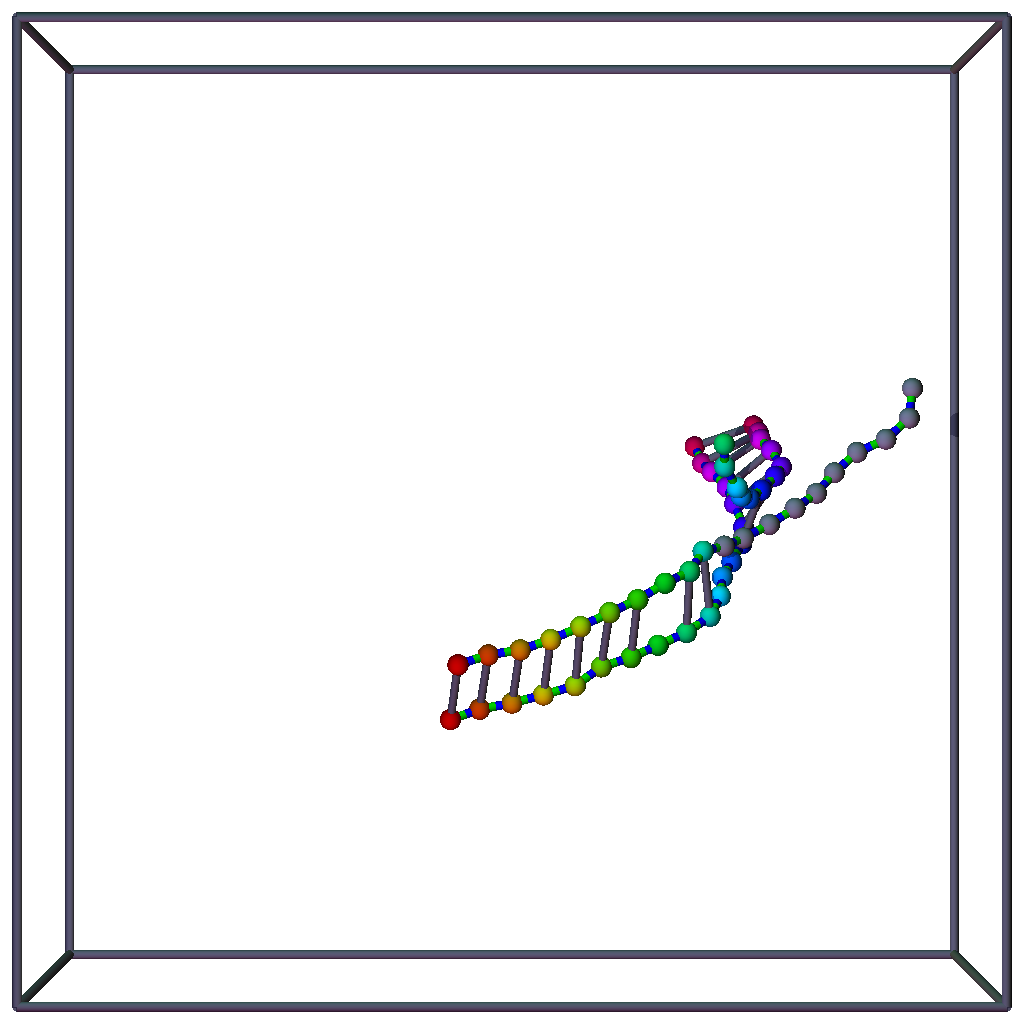}%
\includegraphics[width=0.2\columnwidth]{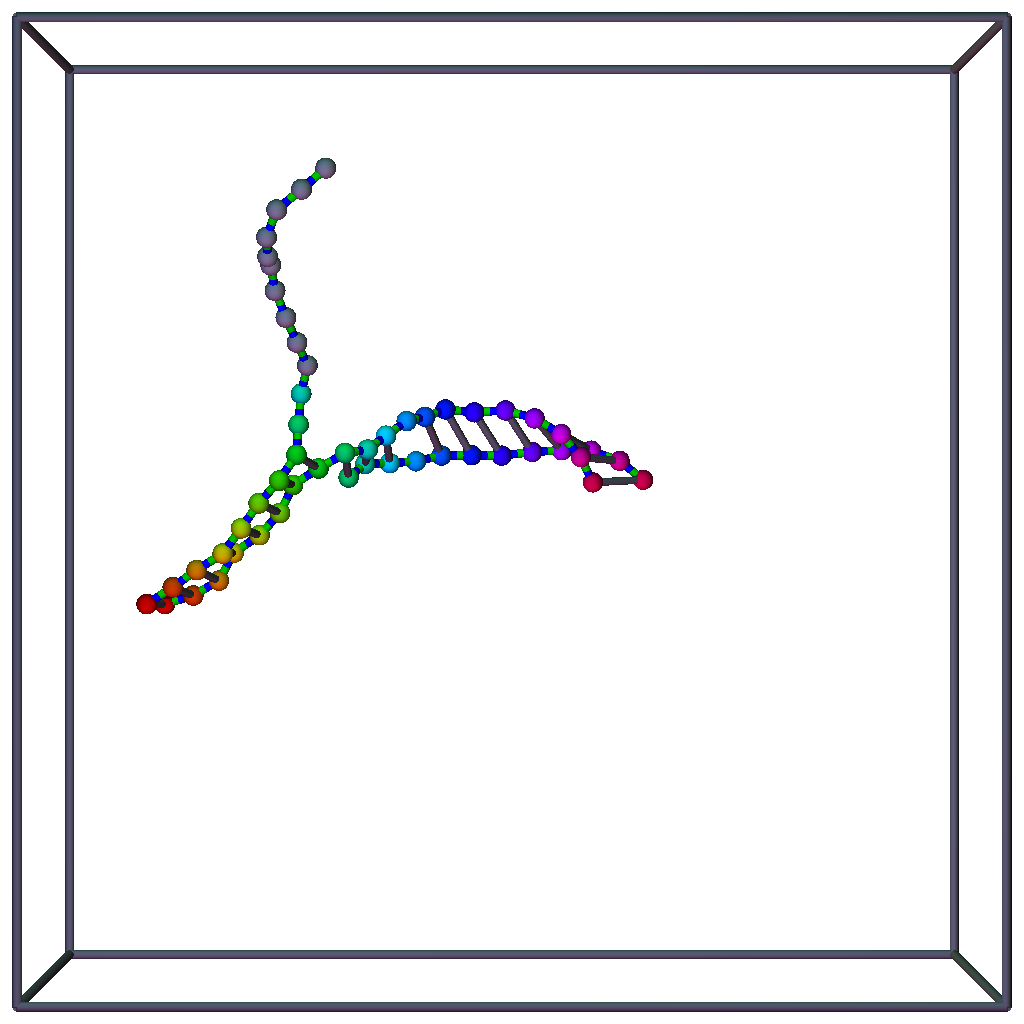}%
\includegraphics[width=0.2\columnwidth]{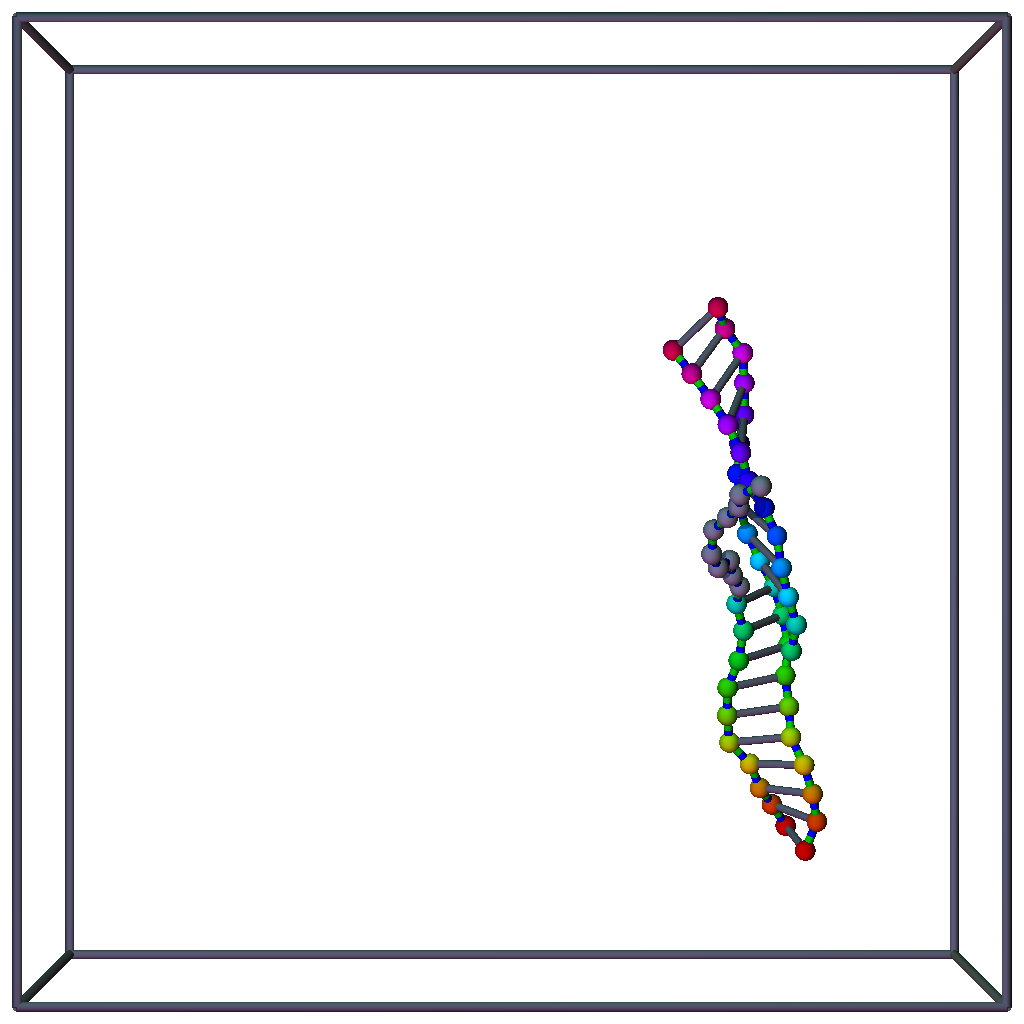}%
\includegraphics[width=0.2\columnwidth]{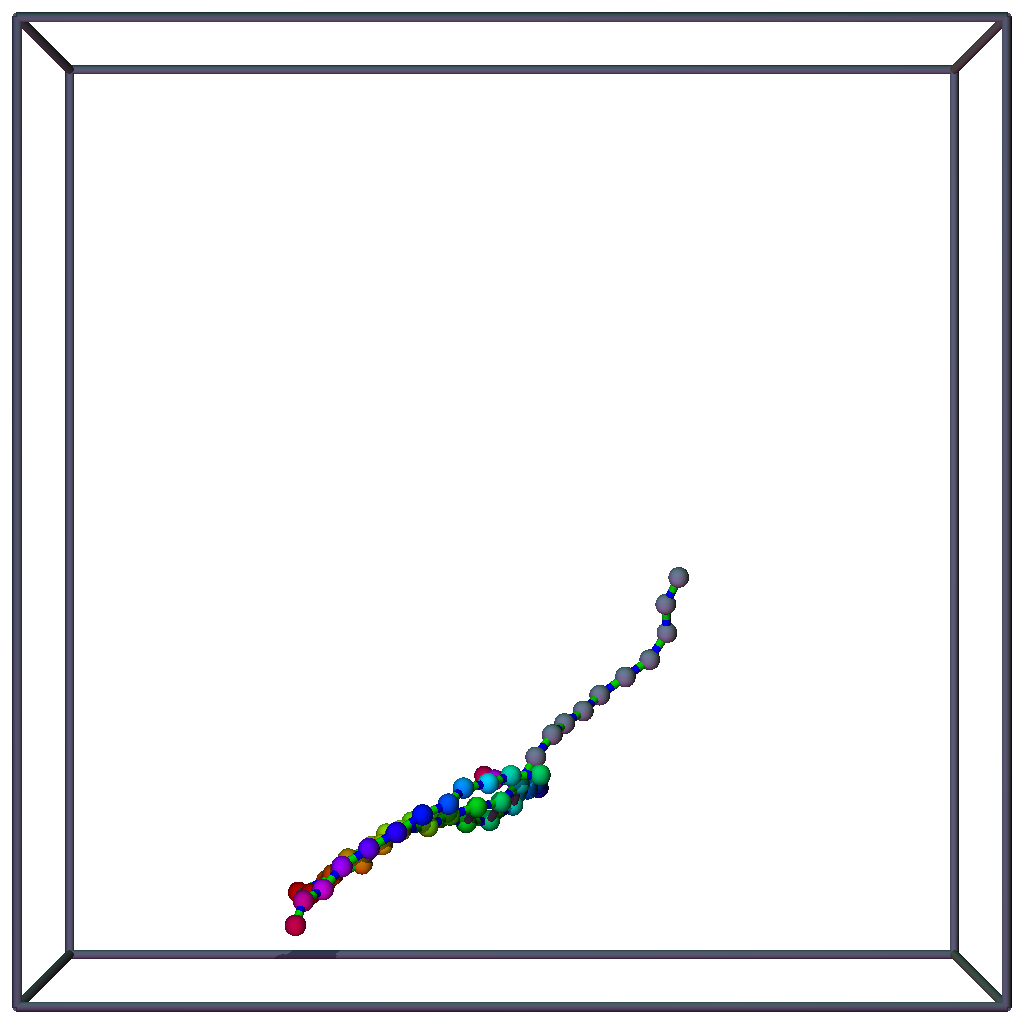}

\caption{\protect\label{fig:Simulations-of-strand-displacement}
Simulations of strand displacement. A $20$ bead long oligomer displaces
a $12$ bead long oligomer initially hybridized to a template for
times $t=$$500$, $1.600$, $1.700$, $1.900\tau_{L}$ (top row).
A $10+10$ bead long oligomer where the latter half is non-complementary
fails to displace a short oligomer hybridized to a template for times
$t=100$, $3.000$, $7.000$, $10.000\tau_{L}$. Simulations are run
at temperature $T=1\epsilon$. Non-complementary beads are show as
gray.}
\end{figure}

\begin{figure}
\centering
  \includegraphics[width=0.8\columnwidth]{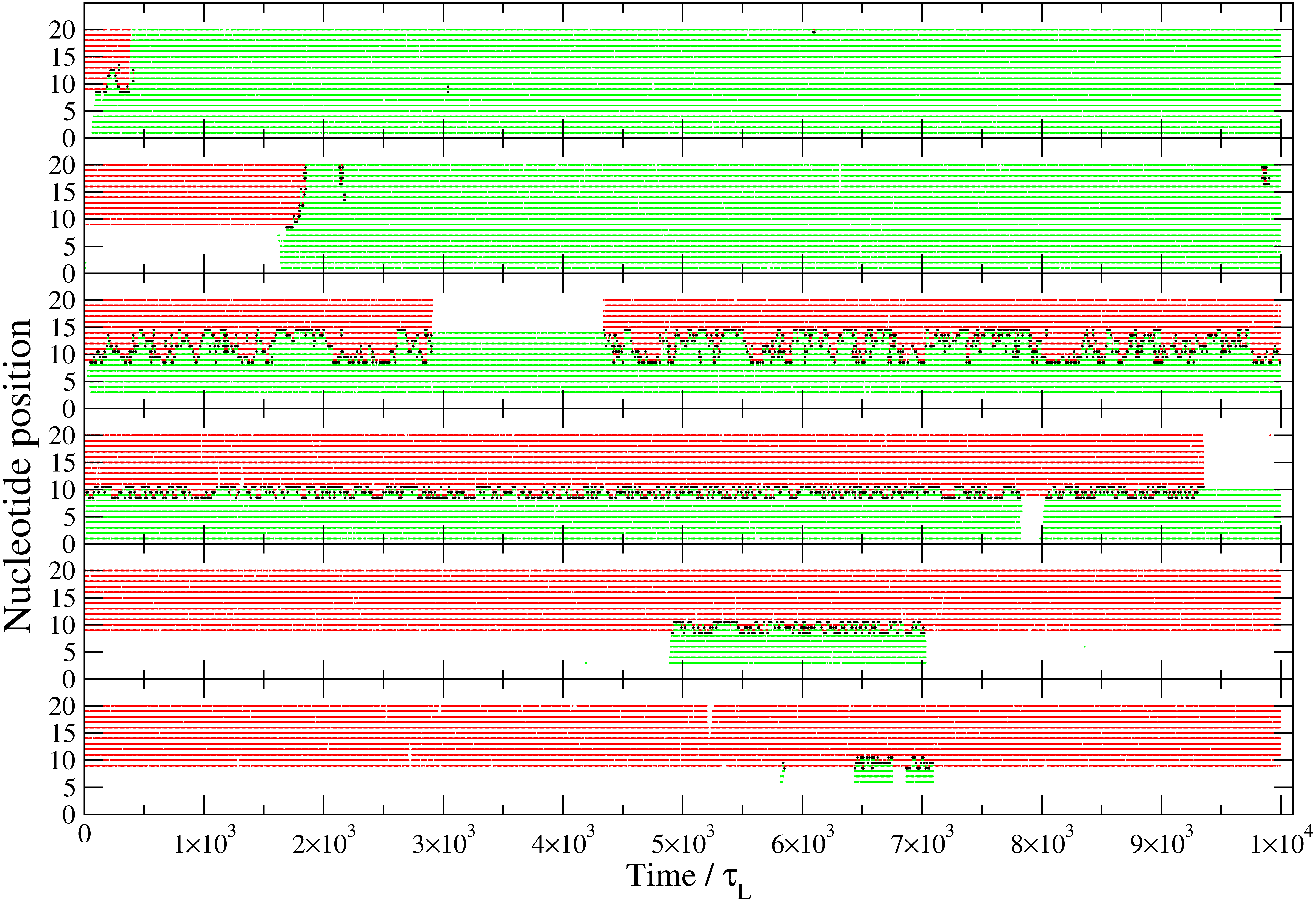}

\caption{\protect\label{fig:Time-evolution}
Time evolution of branch migration on individual template
nucleotide beads. From top to bottom a) A $12$ bead long oligomer (red)
being displaced by a $20$ bead long oligomer (green) on the template. b)
a $20$ bead long oligomer with different random conditions. c) $12$ bead
oligomer (red) competing with a $12+8$ oligomer with 12 complementary
and 8 non-complementary beads (green), d) a $12$ bead oligomer (red)
competing with a $10+10$ oligomer, e) a $12$ bead oligomer (red)
competing with a $8+10$ oligomer, f) a $12$ bead oligomer (red)
competing with  $5+10$ oligomer. Also shown is the branch migration point (black).}
\end{figure}

Fig.~\ref{fig:Simulations-of-strand-displacement} shows simulations
of the strand displacement process underlying Seelig et al.'s DNA
computing approach~\cite{See:06}. The top row shows the successful
displacement of an initially hybridized 12 bead long signal strand
from a 20 bead long template by a 20 bead long signal strand: once
the signal strand diffuses to and binds to the toehold region, branch
migration occurs quickly (during 300 time units) and the formerly
bound signal strand is displaced irreversibly. The bottom row, on
the other hand, shows how the displacement stalls in the presence
of mismatches: here, a mismatch in the domain (last 10 beads) permits
further hybridization of the signal strand. The newly binding and
the original signal strand compete for matching bases in a random
walk process until the non-matching strand dehybridizes again and
leaves the gate available for potential matching signals (not present
in the simulation).

Fig.~\ref{fig:Time-evolution} shows statistics of the displacement
processes for several runs: the graphs depict hybridized bases of
the original (red) and the newly binding signal strand (green), as
well as the branch migration point (black). In the case of matching
signals (top two simulations), it can be seen that displacement occurs
quickly and essentially irreversible once the original strand is fully
displaced. In the third simulation the signal strand and hybridized
strand has the same length, and the interface is seen to diffuse forwards
and backwards. A single dehybridization event is also observed for
the original strand. In the case of mismatching signals (bottom three
simulations), the displacement cannot proceed further than nucleotide
10, and the interface randomly moves between positions 8 and 10, until
-- occasionally -- the mismatching signal dehybridizes from the toehold
region (lack of green markers). In this case, the number of beads
complementary to the toehold region (here 10, 8, and 5 beads) determines
the equilibrium between hybridized and dehybridized configurations,
and thus the performance and availability of the gate. Fig.~\ref{fig:Time-evolution}
also depicts a source of potential failure in logical gates based
on strand displacement, as the output signal can spontaneously dehybridize
even in the absence of a matching input signal (as observed in the
fourth simulation).

\section{Conclusions\label{sec:Conclusions}}

With these initial simulations, we have demonstrated that our coarse-grained
DNA model can succesfully simulate DNA assembly as well as DNA strand displacement
dynamics which form the basis of state-of-the-art DNA computing approaches.
We have successfully simulated self-assembly of DNA tetrahedra and icosahedra
from four and twelve branched DNA constructs, respectively. Simulations show
that the constructs self-assemble into the expected target structures. 

We have further simulated successful displacement of an output strand when
a matching input strand is present. In the presence of mismatches, we could
demonstrate how the displacement process is prevented. Our simulations also
capture potential failures of gates based on strand displacement, namely
spontaneous release of the output strand in the absence of an input signal.
These proof of concept simulations demonstrate how our coarse-grained model
can be used to optimize the length and arrangement of toehold and domain
structures in DNA computing approaches.

While such gate optimizations do not necessarily require spatially resolved
models, our coarse-grained DNA model enables us to study systems that integrate
DNA assembly and computing within a single framework. This enables us to
use these simulations as a starting point for building and testing statistical
mechanical theories describing these complex systems.

\section{Acknowledgements}

The research leading to these results has received funding from the
European Community's Seventh Framework Programme (FP7/2007-2013) under
grant agreement no 249032 (MATCHIT). Funding for this work is provided
in part by the Danish National Research Foundation through the Center
for Fundamental Living Technology (FLinT). 

\bibliographystyle{splncs03}

\begin{thebibliography}{10}
\providecommand{\url}[1]{\texttt{#1}}
\providecommand{\urlprefix}{URL }

\bibitem{Adl:1994}
Adleman, L.M.: Molecular computation of solutions to combinatorial problem.
  Science  266,  1021 (1994)

\bibitem{Andersen2008}
Andersen, E.S., Dong, M., Nielsen, M.M., Jahn, K., Subramani, R., Mamdouh, W.,
  Golas, M.M., Sander, B., H., S., Oliveira, C.L.P., Pedersen, J.S., Birkedal,
  V., Besenbacher, F., Gothelf, K.V., Kjems, J.: Self-assembly of a nanoscale
  {DNA} box with a controllable lid. Nature  459, ~73 (2008)

\bibitem{icosahedra}
Bhatia, D., Mehtab, S., Krishnan, R., Indi, S.S., Basu, A., Krishnan, Y.:
  Icosahedral {DNA} nanocapsules by modular assembly. Angew. Chem.  121,  4198
  (2009)

\bibitem{CHARMM2009}
Brooks, B.R., Brooks, III, C.L., Mackerell, Jr., A.D., Nilsson, L., Petrella,
  R.J., Roux, B., Won, Y., Archontis, G., Bartels, C., Boresch, S., Caflisch,
  A., Caves, L., Cui, Q., Dinner, A.R., Feig, M., Fischer, S., Gao, J.,
  Hodoscek, M., Im, W., Kuczera, K., Lazaridis, T., Ma, J., Ovchinnikov, V.,
  Paci, E., Pastor, R.W., Post, C.B., Pu, J.Z., Schaefer, M., Tidor, B.,
  Venable, R.M., Woodcock, H.L., Wu, X., Yang, W., York, D.M., Karplus, M.:
  {{CHARMM}: The Biomolecular Simulation Program}. {J. Comput. Phys.}  {30},
  {1545} ({2009})

\bibitem{Car:2011}
Cardelli, L.: Strand algebras for {DNA} computing. Nat. Comput.  10,  407
  (2011)

\bibitem{case2010amber}
Case, D.A., Darden, T.A., Cheatham~III, T.E., Simmerling, C.L., Wang, J., Duke,
  R.E., Luo, R., Walker, R.C., Zhang, W., Merz, K.M., et~al.: {AMBER} 11.
  University of California, San Francisco  (2010)

\bibitem{cheatham2000molecular}
Cheatham~III, T.E., Young, M.A.: Molecular dynamics simulation of nucleic
  acids: Successes, limitations, and promise. Biopolymers  56,  232 (2000)

\bibitem{Seeman1991}
Chen, J., Seeman, N.C.: Synthesis from {DNA} of a molecule with the
  connectivity of a cube. Nature  350,  631 (1991)

\bibitem{Erben2007}
Erben, C.M., Goodman, R.P., Turberfield, A.J.: A self-assembled {DNA}
  bipyramid. J. Am. Chem. Soc.  129,  6992 (2007)

\bibitem{Goodman09122005}
Goodman, R.P., Schaap, I.A.T., Tardin, C.F., Erben, C.M., Berry, R.M., Schmidt,
  C.F., Turberfield, A.J.: Rapid chiral assembly of rigid {DNA} building blocks
  for molecular nanofabrication. Science  310,  1661 (2005)

\bibitem{hsu2010theoretical}
Hsu, C.W., Sciortino, F., Starr, F.W.: Theoretical description of a
  {DNA}-linked nanoparticle self-assembly. Phys. Rev. Lett.  105,  55502 (2010)

\bibitem{EveraersKumarSimm2007}
Jost, D., Everaers, R.: A unified description of poly- and oligonucleotide
  {DNA} melting: nearest-neighbor, {P}oland-{S}heraga and lattice models. Phys.
  Rev. E  75,  041918 (2007)

\bibitem{jost2009unified}
Jost, D., Everaers, R.: A unified {P}oland-{S}cheraga model of oligo-and
  polynucleotide {DNA} melting: Salt effects and predictive power. Biophys. J.
  96,  1056 (2009)

\bibitem{JostEveraers2010}
Jost, D., Everaers, R.: Prediction of {RNA} multi-loop and pseudoknot
  conformations from a lattice-based, coarse-grain tertiary structure model. J.
  Chem. Phys.  132,  095101 (2010)

\bibitem{Lak:2012}
Lakin, M.R., Youssef, S., Cardelli, L., Phillips, A.: Abstractions for {DNA}
  circuit design. J R Soc Interface  9,  470 (2012)

\bibitem{langowski2006polymer}
Langowski, J.: Polymer chain models of {DNA} and chromatin. Eur. Phys. J. E
  Soft Matter  19,  241 (2006)

\bibitem{mackerell2000development}
MacKerell~Jr, A.D., Banavali, N., Foloppe, N.: Development and current status
  of the {CHARMM} force field for nucleic acids. Biopolymers  56,  257 (2000)

\bibitem{Martines-VeracoecheaPRL2011}
Martinez-Veracoechea, F.J., Mladek, B.M., Tkachenko, A.V., Frenkel, D.: Design
  rule for colloidcal crystals of {DNA}-functionalized particles. Phys. Rev.
  Lett.  107,  045902 (2011)

\bibitem{ouldridge2010dna}
Ouldridge, T.E., Louis, A.A., Doye, J.P.K.: {DNA} nanotweezers studied with a
  coarse-grained model of {DNA}. Phys. Rev. Lett.  104,  178101 (2010)

\bibitem{OulridgeLouisDoye2011}
Ouldridge, T.E., Louis, A.A., Doye, J.P.K.: Structural, mechanical, and
  thermodynamic properties of a coarse-grained {DNA} model. J. Chem. Phys.
  134,  085101 (2011)

\bibitem{de2011polymer}
de~Pablo, J.J.: Polymer simulations: From {DNA} to composites. Annu. Rev. Phys.
  Chem.  62 (2011)

\bibitem{peyrard2008modelling}
Peyrard, M., Cuesta-Lopez, S., James, G.: Modelling {DNA} at the mesoscale: a
  challenge for nonlinear science? Nonlinearity  21,  T91 (2008)

\bibitem{Lammps}
Plimpton, S.: Fast parallel algorithms for short-range molecular dynamics. J.
  Comp. Phys.  117, ~1 (1995), \url{http://lammps.sandia.gov}

\bibitem{poland1966phase}
Poland, D., Scheraga, H.A.: Phase transitions in one dimension and the
  helix-coil transition in polyamino acids. J. Chem. Phys.  45,  1456 (1966)

\bibitem{Qia:2011}
Qian, L., Winfree, E.: Scaling up digital circuit computation with {DNA} strand
  displacement cascades. Science  332,  1196 (2011)

\bibitem{Rothemund2006}
Rothemund, P.W.K.: Folding {DNA} to create nanoscale shapes and patterns.
  Nature  440,  297 (2006)

\bibitem{sambriski2009sequence}
Sambriski, E.J., Ortiz, V., de~Pablo, J.J.: Sequence effects in the melting and
  renaturation of short {DNA} oligonucleotides: structure and mechanistic
  pathways. J. Phys. Condens. Matter.  21,  034105 (2009)

\bibitem{Sambriski2009}
Sambriski, E.J., Schwartz, D.C., de~Pablo, J.J.: A mesoscale model of {DNA} and
  its renaturation. Biophys. J.  96,  1675 (2009)

\bibitem{SantaLucia}
SantaLucia, J.J., Hicks, D.: The thermodynamics of {DNA} structural motiefs.
  Annu. Rev. Biophys. Biomol. Struct.  33,  415 (2004)

\bibitem{SavelyevPapoian2011}
Savelyev, A., Papoian, G.A.: Chemically accurate coarse graining of
  double-stranded {DNA}. PNAS  107,  20340 (2010)

\bibitem{See:06}
Seelig, G., Soloveichik, D., Zhang, D.Y., Winfree, E.: Enzyme-free nucleic acid
  logic circuits. Science  314,  1585 (2006)

\bibitem{Seeman1982}
Seeman, N.C.: Nucleic acid junctions and lattices. J. Theor. Biol.  99,  237
  (1982)

\bibitem{SvaneborgCPC2012}
Svaneborg, C.: {LAMMPS} framework for dynamic bonding an application modeling
  {DNA}. Comp. Phys. Comm.  183,  1793 (2012)

\bibitem{TinlandMM1997}
Tinlan, B., Pluen, A., Sturm, J., Weill, G.: Persistence length of
  single-stranded {DNA}. Macromolecules  30,  5763 (1997)

\bibitem{Winfree1998}
Winfree, E., Liu, F., Wenzler, L.A., Seeman, N.C.: Design and self-assembly of
  two-dimensional {DNA} crystals. Nature  394,  529 (1998)

\bibitem{Seeman1994}
Xhang, Y., Seeman, N.C.: Construction of a {DNA}-truncated octahedron. J. Am.
  Chem. Soc.  116,  1661 (1994)

\bibitem{Zha:2009}
Zhang, D.Y., Winfree, E.: Control of {DNA} strand displacement kinetics using
  toehold exchange. J. Am. Chem. Soc.  131,  17303 (2009)

\end{thebibliography}

\end{document}